\documentclass[aps,pra,superscriptaddress,longbibliography,twocolumn]{revtex4-2}

\usepackage{graphicx}
\usepackage{amsmath}
\usepackage{amssymb}
\usepackage{hyperref}
\usepackage[utf8]{inputenc}
\usepackage{mathtools}
\usepackage[english]{babel}
\usepackage{bbm}
\hypersetup{colorlinks=true, linkcolor=blue, citecolor=blue, urlcolor=blue}
\usepackage{xcolor}
\usepackage{layouts}
\usepackage{braket}
\usepackage[normalem]{ulem}
\usepackage{mathrsfs}

\newcommand{\ie}{i.\,e.,\ }

\newcommand{\re}{\mathrm{Re}}
\newcommand{\im}{\mathrm{Im}}

\newcommand{\Tr}{\operatorname{Tr}}

\newcommand{\fref}[1]{\text{Fig.}~\ref{#1}}
\newcommand{\ffref}[1]{\text{Figs.}~\ref{#1}}
\newcommand{\eref}[1]{\text{Eq.}~\eqref{#1}}
\newcommand{\eeref}[1]{\text{Eqs.}~\eqref{#1}}

\begin{document}
\title{Breakdown of steady-state superradiance in extended driven atomic arrays}

\author{Stefan Ostermann}
\email{stefanostermann@g.harvard.edu}
\affiliation{Department of Physics, Harvard University, Cambridge, Massachusetts 02138, USA}
\author{Oriol Rubies-Bigorda}
\affiliation{Physics Department, Massachusetts Institute of Technology, Cambridge, Massachusetts 02139, USA}
\affiliation{Department of Physics, Harvard University, Cambridge, Massachusetts 02138, USA}
\author{Victoria Zhang}
\affiliation{Department of Physics, Harvard University, Cambridge, Massachusetts 02138, USA}
\author{Susanne F. Yelin}
\affiliation{Department of Physics, Harvard University, Cambridge, Massachusetts 02138, USA}

\begin{abstract}
Recent advances in generating well controlled dense arrangements of individual atoms in free space have generated interest in understanding how the extended nature of these systems influences superradiance phenomena. Here, we provide an in-depth analysis on how space-dependent light-shifts and decay rates induced by dipole-dipole interactions modify the steady-state properties of coherently driven arrays of quantum emitters. We characterize the steady-state phase diagram, with particular focus on the radiative properties in the steady-state.  Interestingly, we find that diverging from the well-established Dicke paradigm of equal all-to-all interactions significantly modifies the emission properties. In particular, the prominent quadratic scaling of the radiated light intensity with particle number in the steady-state --- a hallmark of steady-state Dicke superradiance --- is entirely suppressed, resulting in only linear scaling with particle number. We show that this breakdown of steady-state superradiance occurs due to the emergence of additional dissipation channels that populate not only superradiant states but also subradiant ones. The additional contribution of subradiant dark states in the dynamics leads to a divergence in the time scales needed to achieve steady-states. Building on this, we further show that measurements taken at finite times for extended atom ensembles reveal properties closely mirroring the idealized Dicke scenario.
\end{abstract}

\maketitle

\section{Introduction}
\label{section: model}
\begin{figure}[t]
    \centering
    \includegraphics{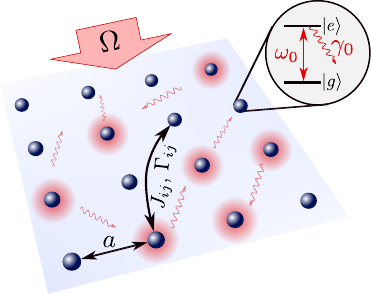}
    \caption{Sketch of the considered model. A dense periodic array of two-level quantum emitters is driven by an external laser with Rabi frequency $\Omega$. If the lattice spacing $a$ is much smaller than the transition wavelength $\lambda_0=2\pi c/\omega_0$, light-induced dipole-dipole interactions will result in non-trivial collective dissipative dynamics.}
    \label{fig:model}
\end{figure}

Dicke superradiance,~\ie the effect where a group of excited atoms collectively emit radiation faster than they would individually, is a fundamental effect in light-matter interacting systems~\cite{gross_haroche}.
Since Dicke introduced the concept of cooperative radiation of atomic ensembles~\cite{Dicke_originalpaper}, a series of seminal works have investigated cooperative phenomena in systems with strong light-matter interaction in various settings. Prominent examples include Dicke superradiance of atomic ensembles coupled to optical resonators and waveguides, in large clouds of Rydberg atoms and dense ensembles of two-level quantum emitters~\cite{exp_super_atoms_1, exp_super_atoms_2, exp_super_atoms_3, exp_super_atoms_4, exp_super_atoms_5, exp_super_atoms_6, exp_super_atoms_7}. More recently the advent of novel technologies which offer more control over the arrangement of individual atoms~\cite{lewenstein_ultracold_2007, bloch_quantum_2012, grier_revolution_2003, barredo_atom-by-atom_2016, endres_atom-by-atom_2016, norcia_microscopic_2018, cooper_alkaline-earth_2018} or quantum emitters~\cite{experiment_quantumdots_1, experiment_quantumdots_2, experiment_NV, experiment_2dmat} has motivated an in-depth study of the role of the emitter geometry on superradiant emission~\cite{masson_many-body_2020, masson_universality_2022, superradiance_cumulants_ours, sierra_dicke_2022, ma_superradiance_2022, mok_dicke_2023, Oriol_superradiance, Robicheaux_superradiance,subradiance_cumulants_ours}. While the dissipative long-range nature of the underlying spin model makes exact studies of these systems challenging, it also provides opportunities to unveil fundamental governing principles of dissipative many-body systems, and to develop applications for novel lasing technologies~\cite{bohnet_steady-state_2012}, the preparation of entangled states~\cite{santos_generating_2022, santos_generation_2023} or the generation of non-classical states of light~\cite{parmee_signatures_2020}.

Most works characterizing superradiance focus on the transient dynamics of fully inverted ensembles. Recently, however, a seminal experimental work investigated the steady-state properties of a dissipative cooperative system under constant classical laser drive~\cite{ferioli_non-equilibrium_2023}. In the Dicke limit, where all atoms are assumed to be located at the same spatial position, and the dissipative interactions among them consequently have the same strength, this system is known to exhibit quadratic scaling of the emitted light intensity with particle number in the steady-state~\cite{gonzalez-tudela_mesoscopic_2013, hannukainen_dissipation-driven_2018, somech_heisenberg-langevin_2023} for sufficiently strong laser drive. For extended ensembles of atoms, geometry dependent energy shifts and decay rates are expected to alter these dynamics.

In this work, we investigate the role of light-induced cooperative shifts and decay rates for steady-state superradiance in driven-dissipative periodic arrays of quantum emitters under strong drive.
While previous work focused on mapping out the steady-state phase diagram of this system~\cite{parmee_signatures_2020}, or perturbative approaches in the weak driving regime for disordered clouds~\cite{bienaime_cooperativity_2013, kupriyanov_mesoscopic_2017}, our focus lies on the full characterization of the superradiant emission properties as a function of the system parameters. The core finding of this work is that going beyond the paradigmatic Dicke limit modifies the radiative steady-state properties significantly. While the radiated light intensity in the steady-state scales quadratically with system size in the Dicke limit, it only scales linearly for extended ensembles (see Sec.~\ref{sec:emission}). For very dense extended ensembles measured at finite times, however, we show that the scaling approaches that of the Dicke limit again, namely $N^2$ (see Sec.~\ref{sec:finite_time}).

\section{Model}
We model the system depicted in~\fref{fig:model}(a) as a set of two-level atoms with ground state $\ket{g_i}$ and excited state $\ket{e_i}$, located at positions $\mathbf{r}_i$ in the $x$-$y$ plane with a transition dipole matrix element $\mathbf{d}$. To reduce the number of tunable parameters throughout this work, we choose $\mathbf{d}=(0,0,1)^T$. Analogous results, however, can be found for other polarizations. The Hamiltonian is then given as (we set $\hbar = 1$ here and for the remainder of this work) $\hat{H} = \hat{H}_\mathrm{int} + \hat{H}_\mathrm{drive}$, with
\begin{subequations}
\begin{align}
    \hat{H}_\mathrm{int} &= -\Delta \sum_{i=1}^{N}\hat{\sigma}^+_i\hat{\sigma}^-_{i} + \sum_{i,j\neq i}^N J_{ij}\hat{\sigma}_i^+ \hat{\sigma}_j^-, \label{eq:dipole_Hamiltonian}\\
    \hat{H}_\mathrm{drive} &= \frac{\Omega}{2}\sum_{i=1}^N\left(e^{i\mathbf{k}\cdot\mathbf{r}_i} \hat{\sigma}_i^+ + e^{-i\mathbf{k}\cdot\mathbf{r}_i} \hat{\sigma}_i^-\right),
\end{align}
\label{eq:Hamiltonian}
\end{subequations}
where $\hat{\sigma}_i^+ = |e_i \rangle \langle g_i |$ ($\hat{\sigma}_i^- = |g_i \rangle \langle e_i |$) is the raising (lowering) operator for atom $i$, $\Delta = \omega - \omega_0$ denotes the detuning between the drive laser frequency $\omega$ and the atomic transition frequency $\omega_0$ (including the Lamb shift $J_{ii}$), and $J_{ij}$ is the coherent interaction strength between distant emitters. For the analysis below we choose $\Delta = 0$ unless stated otherwise. The ensemble is driven with a plane wave drive  with constant Rabi frequency $\Omega$ and wavevector $\mathbf{k}$. The direction of $\mathbf{k}$,~\ie the direction of the incoming pump beam, does not alter the steady-state properties, but it affects the transient dynamics and therefore is an important quantity to consider when analyzing finite time measurements in Sec.~\ref{sec:finite_time}. The strength of the coherent interaction is determined via $J_{ij}  = -\frac{3\pi \gamma_0}{\omega_0} {\mathbf{d}^\dagger} \cdot \re\left[\textbf{G}(\textbf{r}_{ij}, \omega_0) \right]\cdot\mathbf{d},$~\cite{Lehmberg_1970_1,Lehmberg_1970_2}, where $\mathbf{G} (\mathbf{r},\omega)$ is the Green's tensor for a point dipole in vacuum~\cite{Chew_dyadicGreens,dyadic_novotny_hecht_2006} given in Appendix~\ref{app:Greensfunction}, and $\mathbf{r}_{ij}=\mathbf{r}_i - \mathbf{r}_j$ is the vector connecting atoms $i$ and $j$.

The dissipative nature of the system is described by a Lindbladian of the form
\begin{equation}
\label{eq: Linbladian}
    \mathcal{L}[\hat{\rho}]  = \! \sum_{i,j=1}^N \! \frac{\Gamma_{ij}}{2} \left( 2 \hat{\sigma}_i^- \hat{\rho} \hat{\sigma}_j^+ - \hat{\sigma}_i^+ \hat{\sigma}_j^- \hat{\rho} - \hat{\rho} \hat{\sigma}_i^+ \hat{\sigma}_j^- \right),
\end{equation}
where $\Gamma_{ij} = \frac{6\pi \gamma_0}{\omega_0} {\mathbf{d}^\dagger} \cdot \im\left[\textbf{G}(\textbf{r}_{ij}, \omega_0)\right] \cdot \mathbf{d}$ describes correlated cooperative decay.
\begin{figure}
    \centering
    \includegraphics{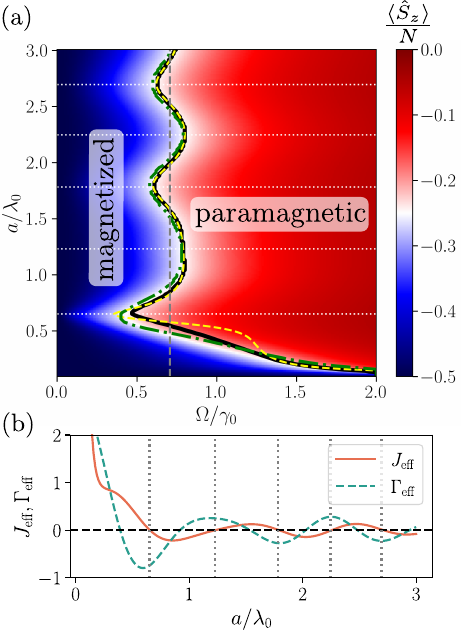}
    \caption{(a) Steady-state phase diagram for a square with the collective spin pointing in $z$-direction, $\langle S_z \rangle/N$, as a function of Rabi frequency $\Omega$ and lattice spacing $a$ for a four atom square. The solid black line indicates the numerically obtained threshold from the full quantum solution, the green dash dotted line indicates the threshold obtained via the mean-field equations, and the dashed yellow line shows the analytically obtained threshold given by Eq.~\eqref{eqn:thres}. The horizontal dashed white lines indicate the lattice spacings at which the effective coupling strengths $J_\mathrm{eff}$ shown in (b) is zero. (b) Effective coupling strength $J_\mathrm{eff}$ and decay rate $\Gamma_\mathrm{eff}$ as a function of lattice spacing. The vertical dashed lines indicate zero crossings of $J_\mathrm{eff}$.}
    \label{fig:phase_diag}
\end{figure}

The full system dynamics is then governed by the master equation for the atomic density matrix $\hat{\rho}$ \cite{Lehmberg_1970_1,Lehmberg_1970_2, CohenTanoudgi_book}
\begin{equation}
\label{eqn:EOM_densitymatrix}
    \frac{d\hat{\rho}}{dt} = -\frac{i}{\hbar} \left[\hat{H} , \hat{\rho}\right] + \mathcal{L}[\hat{\rho}].
\end{equation}
Notably, the resulting long-range interacting spin model is not integrable, and the size of the density matrix $\hat{\rho}$ scales exponentially in system size. While a truncation of the Hilbert space into the so-called single excitation subspace is feasible in the weak driving regime, studying the effects of a strong drive requires to take into account the full exponential Hilbert space. This makes an exact quantum mechanical treatment of this many-body problem impossible for large system sizes, and has traditionally rendered the study of dipole-coupled ensembles in the strong driving regime unfeasible. To circumvent this limitation, we use approximative tools (mean-field approximation and second order cumulant expansions~\cite{Cumulant_Kubo, Ritsch_cumulants_package, Ritsch_cumulants_dipole, robicheaux_intensity_2023} in the analysis below to model larger systems and justify our findings by providing intuition based on the full quantum model in~\eref{eqn:EOM_densitymatrix} for small atom numbers.

An important limit of this model is the so-called Dicke limit, which corresponds to $N$ indistiguishable atoms located at a single position, for example $\mathbf{r}_i = 0$. The governing equations can be obtained from Eqs.~\eqref{eq:dipole_Hamiltonian} and~\eqref{eq: Linbladian} by setting $J_{ij} = 0$ and $\Gamma_{ij} = \gamma_0$, as well as defining collective raising and lowering operators $\hat{S}^\pm = \sum_{i=1}^N \hat{\sigma}^\pm_i$. If the system is initialized in the ground state, the collective dynamics is restricted to the fully symmetric spin sector $\ket{S=N/2, m=-N/2...N/2}$, often also referred to as the Dicke ladder. The Hamiltonian for the coherent drive then reduces to $H_\mathrm{drive}^\mathrm{Dicke} = \Omega/2 (\hat{S}^+ + \hat{S}^-)$ and the Lindbladian can be written as $\mathcal{L}^\mathrm{Dicke}[\rho] = \gamma_0/2 ( 2 \hat{S}^- \hat{\rho} \hat{S}^+ - \hat{S}^+ \hat{S}^- \hat{\rho} - \hat{\rho} \hat{S}^+ \hat{S}^- )$. The fact that the restriction to the symmetric subsector no longer holds in the free space case for finite lattice spacings is pivotal for the results discussed below.

\section{Steady-State Phase Diagram}
\begin{figure*}
    \centering
    \includegraphics[width = 0.95\textwidth]{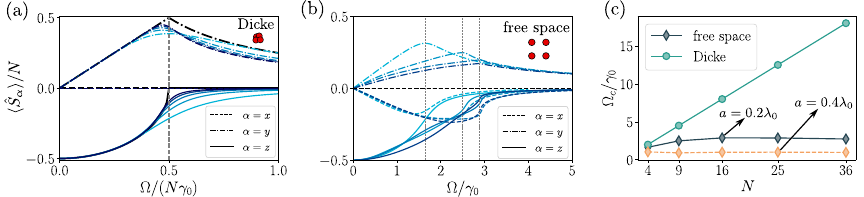}
    \caption{(a) steady-state value of the collective spin expectation value $\langle \hat{S}_\alpha\rangle$ ($\alpha\in\{x,y,z\}$) obtained by solving the master equation~\eqref{eqn:EOM_densitymatrix} for the steady-state in the Dicke limit as a function of driving strength $\Omega$. The thick black lines show the analytical solution obtained from a mean-field model (see Appendix~\ref{app:Dicke_thres}). (b) steady-state solutions of the mean-field model given in~\eref{eqn:full_mean-field} for a square lattice in free space with lattice spacing $a=0.2\lambda_0$. The different shadings in (a) and (b) indicate different particle numbers $N=4,36,64,100$ from light to dark. In either case, we use the maximum of $\langle S_y\rangle$ as the threshold condition.  (c) Scaling of the critical driving strength $\Omega_c$ as a function of particle number. As suggested by (a) and (b), the critical driving strength scales linearly with particle number in the Dicke limit, but it doesn't scale with $N$ for the free space case. This feature is independent of lattice spacing, as  evinced by the solid grey line ($a=0.2\lambda_0$) and the orange dashed line ($a=0.4\lambda_0$).
    }
    \label{fig:mean-field_cuts}
\end{figure*}
We first determine the steady-state phase diagram as a function of $\Omega$ and $a$. It is convenient to express the Hamiltonian $\hat{H}$ in the Pauli basis consisting of the Pauli matrices $\hat{\sigma}_{x,y,z}$ by employing the relation $\hat{\sigma}^{\pm}_i = (\hat{\sigma}_x^i \pm  i \hat{\sigma}_y^i)/2$. Then, the definition of the collective spin operators $\hat{S}_{x,y,z} = 1/2 \sum_{i=1}^N \hat{\sigma}_{x,y,z}$ allows the interpretation of the results based on a collective spin on a single Bloch sphere. An exemplary phase diagram for a square consisting of four atoms is shown in~\fref{fig:phase_diag}(a). It exhibits two distinct regions indicated in blue and red. The blue region corresponds to the trivial magnetized phase where all the spins are polarized, whereas the red region marks the paramagnetic phase where $\langle \hat{S}_z \rangle \approx 0$. Note that related characterizations of the superradiant phase transition in the Dicke limit were presented in Refs.~\cite{gonzalez-tudela_mesoscopic_2013, hannukainen_dissipation-driven_2018, somech_heisenberg-langevin_2023, ferioli_non-equilibrium_2023}. In the Dicke limit, the saturated paramagnetic regime also corresponds to the regime exhibiting enhanced superradiant emission, as will be further discussed in Sec.~\ref{sec:emission}. In the Dicke case, the transition between the magnetized and the paramagnetic phase can be inferred by a peak in the value for the $y$ component of the collective spin $\langle \hat{S}_y\rangle$ [see~\fref{fig:mean-field_cuts}(a)]. In the finite size free space case, this peak is no longer as pronounced [see~\fref{fig:mean-field_cuts}(b)]. Nevertheless, the maximum of $\langle \hat{S}_y\rangle$ remains a good indicator for the transition point between the magnetized and paramagnetic phase. Hence, we use $\max\{\langle \hat{S}_y \rangle\}$ to characterize the transition between these two regimes for the remainder of this work.

To study the particle number scaling of this superradiant transition and to obtain some analytical insights, we first focus on a mean-field model. The equations of motion for the expectation values $s^{x,y,z} \equiv \langle \hat{\sigma}_{x,y,z}\rangle$ can be obtained from~\eref{eqn:EOM_densitymatrix} via the expression $\partial_t\langle\hat{O}\rangle = \Tr([\partial_t \hat{\rho}(t)] \hat{O})$, and by performing a mean-field approximation for the two-point correlators $\langle \hat{A}\hat{B}\rangle = \langle \hat{A}\rangle \langle\hat{B} \rangle$. This results in a set of $3N$ coupled differential equations governing the dynamics of the form
\begin{subequations}
\begin{align}
\dot{s}_k^x &= \sum_{i \neq k} J_{ki} s_i^y s_k^z - \frac{\gamma_0}{2} s_k^x +\sum_{i\neq k}\frac{\Gamma_{ki}}{2} s_i^x s_k^z, \\
\dot{s}_k^y &= -\sum_{i \neq k} J_{ki} s_i^x s_k^z - \frac{\gamma_0}{2} s_k^y +\sum_{i\neq k}\frac{\Gamma_{ki}}{2} s_i^y s_k^z - \Omega s_k^z, \\
\dot{s}_k^z &= \sum_{i \neq k} J_{ki} (s_k^x s_i^y - s_i^y s_k^x) - \gamma_0(1+s_k^z) \nonumber\\
& \quad \,- \sum_{i\neq k}\frac{\Gamma_{ki}}{2} (s_k^x s_i^x + s_i^y s_k^y) + \Omega s_k^y,
\end{align}
\label{eqn:full_mean-field}
\end{subequations}
for $i$ and $k$ $\in\{1...N\}$ and pump perpendicular to the $x$-$y$ plane, such that the atom-dependent driving phase in~\eref{eq:Hamiltonian} vanishes ($\mathbf{k}\cdot \mathbf{r}_i=0$ for $i=1,...,N$) (see Appendix~\ref{app:cumulants} for the more general set of equations). These equations capture the transition between the paramagnetic and magnetized phase remarkably well. In particular, the threshold obtained from the mean-field equations matches with that obtained with the full master equation~\eqref{eqn:EOM_densitymatrix}, which are respectively shown by the dash-dotted green trace and the solid black trace in~\fref{fig:phase_diag}(a).
Performing an additional approximation allows us to obtain analytical insights based on this mean-field model. For infinite periodic arrays, we can leverage the geometries' symmetry and define the effective interaction strength and decay rate as $J_\mathrm{eff} = \sum_{i=2}^N J_{1 i}$ and $\Gamma_\mathrm{eff} = \sum_{i=2}^N \Gamma_{1 i}$, respectively. For subwavelength lattices these effective couplings converge to finite constants as $N \rightarrow \infty$, for which the approximation becomes exact.
This simplifies the set of equations to just three equations , which describe the mean-field dynamics of a single spin surrounded by all other atoms. They are given as
\begin{subequations}
\label{eqn:mf_simple}
\begin{align}
\dot{s}^x &= J_\mathrm{eff} s^y s^z - \frac{1}{2}(\gamma_0 - \Gamma_\mathrm{eff} s^z)s^x,\\
\dot{s}^y &= -J_\mathrm{eff} s^x s^z - \frac{1}{2}(\gamma_0 - \Gamma_\mathrm{eff} s^z)s^y - \Omega s^z,\\
\dot{s}^z &= -\gamma_0(1+s^z) - \frac{1}{2}\Gamma_\mathrm{eff}\left[(s_x)^2 + (s_y)^2\right] + \Omega s^y.
\end{align}
\end{subequations}
Linearizing this set of equations around the steady-state solution via $s_{x,y,z} = s_0^{x,y,z} + \delta s_{x,y,z}(t)$ and solving for the steady-state in $\delta s_{x,y,z}(t)$ (see Appendix~\ref{app:lin_eq}) allows the extraction of a concise expression for the critical driving strength. In accordance with the Dicke limit [see~\fref{fig:mean-field_cuts}(a)], we again define the value of $\Omega$ where $s^y$ is maximal as the critical driving strength $\Omega_c$. We obtain (see Appendix~\ref{app:lin_eq})
\begin{equation}
\Omega_c = \frac{\sqrt{\gamma_0}\sqrt{4 J_\mathrm{eff}^2 + (\gamma_0 +\Gamma_\mathrm{eff})^2}}{\sqrt{2}\sqrt{\gamma_0 + \Gamma_\mathrm{eff}}}.
\label{eqn:thres}
\end{equation}
 Despite the substantial approximation performed with the linearization of equations~\eqref{eqn:mf_simple}, this analytical treatment captures the overall properties of the phase diagram, as can be seen from the yellow dashed line in~\fref{fig:phase_diag}(a). In particular, $\Omega_c$ has extrema at lattice spacings where $J_\mathrm{eff} = 0$ (see~\fref{fig:phase_diag}). These extrema coorespond to a maximum (minimum) when $\Gamma_\mathrm{eff} > 0$ ($\Gamma_\mathrm{eff} < 0$).

The critical driving strength given in~\eref{eqn:thres} is constant in the particle number $N$. This marks a strong difference to the Dicke case, where the critical driving strength in the thermodynamic limit, $\Omega_c^\mathrm{Dicke} = N \gamma / 2$ (see Appendix~\ref{app:Dicke_thres}), scales linearly with particle number. This distinction between the free space and the Dicke case is illustrated in~\fref{fig:mean-field_cuts}. The full quantum solution of the Dicke model approaches the analytical steady-state solution (see Appendix~\ref{app:Dicke_thres}) for $N\rightarrow \infty$. For the free space case, we rely on the mean field model in~\eref{eqn:full_mean-field} to obtain the particle number scaling. Intriguingly, the critical value $\Omega_c$ does not scale with particle number, except for finite size effects at very small system sizes[see~\fref{fig:mean-field_cuts}(b)]. The different scaling of the critical driving strength in particle number suggests that the two cases studied here are part of different universality classes. While a detailed analysis of the models' criticality in the thermodynamic limit warrants further study, it goes beyond the scope of the present work.

\section{Emission Properties}
\label{sec:emission}
\begin{figure}
    \centering
    \includegraphics{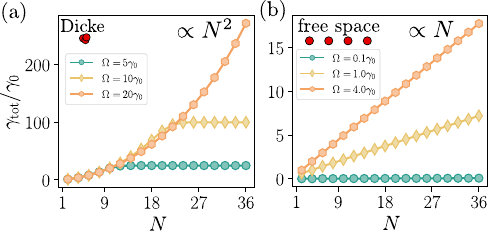}
    \caption{Scaling of the total emission rate as a function of particle number (a) for the Dicke case, and (b) for a chain with lattice spacing $a=0.2\lambda_0$ in free space, obtained by evolving the equations in second order cumulant expansion until a steady-state is reached. The different colors and markers indicate different driving strengths. For sufficiently strong driving, we find a quadratic scaling $\propto N^2$ with particle number for the Dicke case and a linear scaling $\propto N$ for the free space case.}
    \label{fig:ss_scaling}
\end{figure}
One of the core properties of steady-state superradiance in the Dicke limit is the quadratic scaling of the total emission rate $\gamma_\mathrm{tot} = \sum_{i j}\Gamma_{i,j}\langle\hat{\sigma}_i^+ \hat{\sigma}_j^-\rangle$ in the steady-state ($\gamma_\mathrm{tot}^\mathrm{ss}$) with particle number $N$~\cite{ferioli_non-equilibrium_2023,somech_heisenberg-langevin_2023}.  This rate is given as the expectation value of either the jump term or the anti-commuting part of the Lindbladian in~\eref{eq: Linbladian} via  $\gamma_\mathrm{tot} = \mathrm{Tr} \left\{  \! \sum_{i,j} \! \frac{\Gamma_{ij}}{2} \left( \hat{\sigma}_i^+ \hat{\sigma}_j^- \hat{\rho} + \hat{\rho} \hat{\sigma}_i^+ \hat{\sigma}_j^- \right) \right\} =  \mathrm{Tr} \left\{  \! \sum_{i,j} \! \Gamma_{ij} \hat{\sigma}_j^- \hat{\rho} \hat{\sigma}_i^+ \right\} = \langle \sum_{i,j} \Gamma_{ij} \hat{\sigma}_i^+ \hat{\sigma}_j^- \rangle \equiv \langle \hat{H}_\mathrm{dis}\rangle$, where we defined the dissipative Hamiltonian $\hat{H}_\mathrm{dis}$. In~\fref{fig:ss_scaling}(a) we plot $\gamma_\mathrm{tot}^\mathrm{ss}$ obtained from a solution of the master equation in the Dicke limit as a function of $N$ for different driving strengths $\Omega$. For sufficiently large $\Omega$, the total emission rate scales quadratically with particle number. For small $\Omega$, the emission rate saturates to a constant value above a certain particle number [see~\fref{fig:ss_scaling}(a)]. This occurs because the decay rates of the states in the symmetric Dicke ladder scale at least with N, and small values of $\Omega$ are not enough to sufficiently invert large systems to attain the $N^2$ scaling characteristic of the $|S=N/2,m=0 \rangle$ state. For subwavelength arrays in free space, where the underlying geometry results in cooperative shifts and decay rates, a natural question arises: How does the geometry influence the emission properties, in particular in the superradiant regime?
\begin{figure*}
    \centering
    \includegraphics[width = 0.9\textwidth]{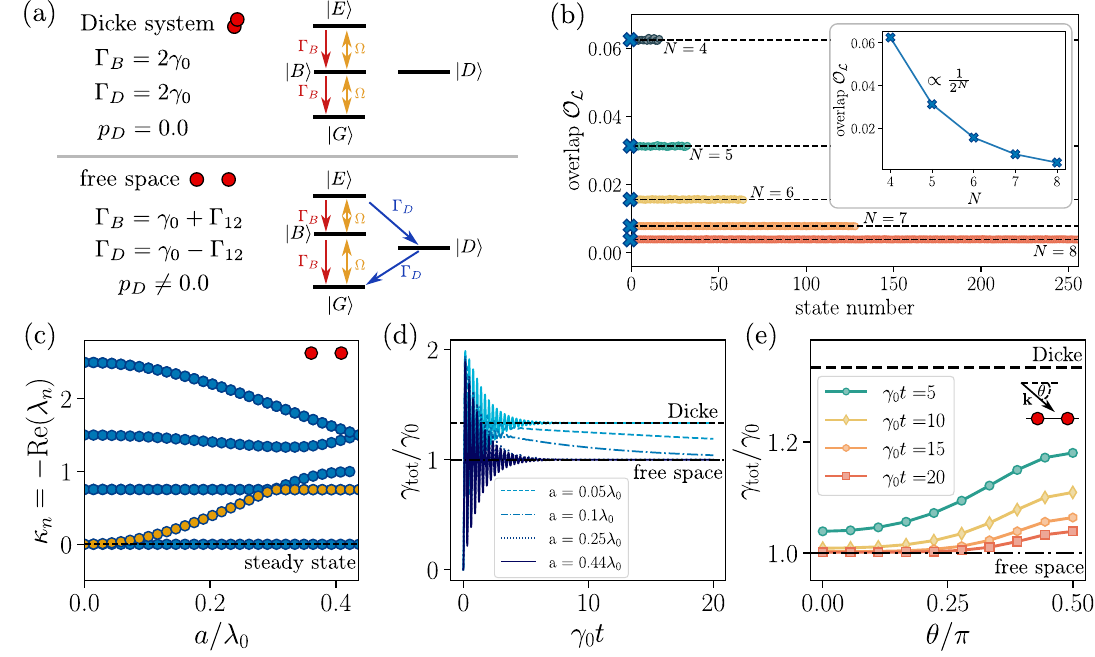}
    \caption{(a) Intuitive explanation for the different scaling of the total emission rate with the particle number $N$ based on the two atom case. In the Dicke case (top), decay from $\ket{E}$ to $\ket{G}$ only occurs via the bright state $\ket{B}$ at a rate $\Gamma_B = 2\gamma_0$. Similarly, the driving field only brings population back into the fully excited state $\ket{E}$ via the symmetric Dicke ladder, \ie via state $\ket{B}$. As a result, the steady-state at large drive contains an equal population in $\ket{G}$, $\ket{B}$ and $\ket{E}$. In the free space case (bottom), decay from $\ket{E}$ to $\ket{G}$ can occur either through the bright state $\ket{B}$ at a rate $\Gamma_B = \gamma_0 + \Gamma_{12}$, or through the dark state $\ket{D}$ at a rate $\Gamma_D = \gamma_0 - \Gamma_{12}$ Consequently, the steady-state exhibits equal population in all four states $\ket{G}$, $\ket{B}$, $\ket{E}$ \emph{and} $\ket{D}$, which results in a diminished total emission rate. This occurs even in the limit $a/\lambda_0\rightarrow 0$, when $\Gamma_{12} \rightarrow \gamma_0$ and $\Gamma_{D} \rightarrow 0$. Notably, the intuition gained from the two atom case also holds for larger particle numbers. (b) Overlap of the steady-state with the full Lindbladian spectrum for a chain with lattice spacing $a=0.2\lambda_0$ and strong pump $\Omega =  40 \gamma_0$ for multiple particle numbers $N$. In all cases, all states of the spectrum are equally populated in the steady-state. (c) Decay rate $\kappa_n = -Re(\lambda_n)$ of the eigenmodes of the Liouvillian in \eref{eqn:EOM_densitymatrix} as a function of spacing $a$. The eigenvector associated with the eigenvalue $\kappa_n=0$ corresponds to the steady-state. The timescale needed to reach this steady-state is equal to the inverse of the smallest non-zero decay rate (orange markers) of the Liouvillian spectrum. For $a/\lambda_0 \rightarrow 0$, this timescale diverges. (d) Emission rate as a function of time for two atoms separate by different distances $a$, illuminated by a plane wave drive perpendicular to the atomic chain. As $a$ decreases, the time required to reach the steady-state increases and the evolution resembles that of the Dicke limit for longer times. (e) Total emission rate measured at different finite times as a function of the angle of the impinging driving field with the atomic chain. Larger emission rates are attained for perpendicular drive, $\theta = \pi/2$, than for drive parallel to the atom chain, $\theta = 0$. A spacing of $a = 0.1 \lambda_0$ is considered.}
    \label{fig:explan_fig}
\end{figure*}

Since solving the full master equation~\eqref{eqn:EOM_densitymatrix} is not feasible in the free space case, we perform a cumulant expansion up to second order~\cite{Cumulant_Kubo, Ritsch_cumulants_package, Ritsch_cumulants_dipole, robicheaux_intensity_2023} to extract the scaling of the total steady-state emission rate. The full set of equations can be found in Appendix~\ref{app:cumulants}. In contrast to the quadratic scaling in the Dicke limit, we find linear scaling with particle number for any value of $\Omega$ in the free space case [see~\fref{fig:ss_scaling}(b)], independent of spacing or geometry. This drastic change in the emission properties arises from the space dependent coherent $J_{ij}$ and dissipative $ \Gamma_{ij}$ dipole interactions, which result in  additional decay channels that couple bright to subradiant states through dissipation. These decay channels and the subsequent coupling to subradiant states are suppressed in the Dicke limit, which is restricted to the symmetric bright or radiating states [see~\fref{fig:explan_fig}(a)]. While the data presented in~\fref{fig:ss_scaling} are for a chain of atoms, qualitatively similar results are obtained for a square array or other higher dimensional geometries.

The general mechanism resulting in this stark deviation from the Dicke limit can be already illustrated in a two atom model. For two atoms at a distance $a<\lambda_0$, the dipole-dipole interactions become significant and give rise to coherent interactions $J_{12}$ and correlated emission or dissipation $\Gamma_{12}$. The eigenstates of the Hamiltonian~\eqref{eq:Hamiltonian} are then given as $\ket{G} = \ket{gg}$, $\ket{B} = 1/\sqrt{2}(\ket{eg} + \ket{ge})$, $\ket{D} = 1/\sqrt{2}(\ket{eg} - \ket{ge})$ and $\ket{E} = \ket{ee}$, where $\ket{B}$ ($\ket{D}$) denotes the symmetric (antisymmetric) single-excitation bright (dark) state.
These eigenstates are coupled via different dissipation channels. While the symmetric bright state $\ket{B}$ is coupled to $\ket{E}$ and $\ket{G}$ via the cooperatively enhanced bright channel with a decay rate $\Gamma_B = \gamma_0 + \Gamma_{12}$, the antisymmetric dark state couples to these states via the suppressed decay rate $\Gamma_D = \gamma_0 - \Gamma_{12}$. In the Dicke limit, the distance between emitters tends to zero and the dissipative interaction approaches the spontaneous decay rate, $\lim_{|\mathbf{r}_{12}|\rightarrow 0} \left(\Gamma_{12}\right) = \gamma_0$~ \cite{masson_universality_2022, Samuel_PRA}. As a result, the dynamics is restricted to the symmetric subspace,~\ie the Dicke ladder $\ket{E} \rightarrow \ket{B} \rightarrow \ket{G}$ [see~\fref{fig:explan_fig}(a)], and the population of the dark state is zero at all times ($p_D^\mathrm{Dicke} = 0$). This changes in the free space case, where an additional dissipation channel to the dark state emerges~\cite{Bloch_mirror} and $\Gamma_D$ takes a small but finite value, $\Gamma_D \neq \gamma_0$.
Then, the decay into the dark state is no longer fully suppressed [see~\fref{fig:explan_fig}(a)], and a finite population in the dark state ($p_D^\mathrm{f.s.} \neq 0$) is attained. For this simple two atom model, the populations of the individual states in the steady-state can be obtained analytically (see Appendix~\ref{app:two-atoms}). In the Dicke case and for sufficiently strong driving on resonance with the bright state,~\ie $\Delta = J_{12}$, the interplay of continuous drive and collective dissipation along the Dicke ladder results in an equillibrium configuration where all states in the symmetric subspace are equally populated,~\ie $p_G^\mathrm{Dicke} = p_B^\mathrm{Dicke} = p_E^\mathrm{Dicke} = 1/3$, whereas the dark state remains unoccupied $p_D^\mathrm{Dicke} = 0$. In the free space case, however, the additional dissipation channel $\propto \Gamma_D$ modifies the equilibrium state, such that \emph{all} four eigenstates are populated equally in the steady-state,~\ie $p_G^\mathrm{f.s.} = p_B^\mathrm{f.s} = p_D^\mathrm{f.s} = p_E^\mathrm{f.s} = 1/4$. That is, a significant amount of the population is then trapped in a non-radiative dark state.  This results in a reduced emission rate in the free space case, and will ultimately lead to the different scaling of the emission properties with particle number shown in Fig.~\ref{fig:ss_scaling}.

Crucially, the steady-state populations at large drive follow the same trend for a general particle number N: in the free space case, all states are equally populated; in the Dicke case, only the states within the symmetric subspace are equally populated. While a full analytical solution for the steady-state of~\eref{eqn:EOM_densitymatrix} is cumbersome in this general setting, we can numerically test this intuition for small atom numbers. In particular, the steady-state solution can be determined via the spectrum $\{\lambda_n = -\kappa_n + i \nu_n \}$ of the Liouvillian $\mathscr{L}[\hat{\rho}] = i/\hbar[\hat{H},\hat{\rho}] + \mathcal{L}[\hat{\rho}]$, where $\kappa_n$ and $\nu_n$ respectively denote the decay rate and energy shift associated to the $n$-th eigenvalue. More precisely, the steady-state fulfills $\partial_t{\hat{\rho}_{ss}} = 0$ and therefore corresponds to the Liouvillian eigenstate with zero decay rate, \ie the state with $\kappa_n=0$. In~\fref{fig:explan_fig}(b), we show the overlap $\mathcal{O}_\mathscr{L} = \bra{\psi}\hat{\rho}_{ss}\ket{\psi}$ of the steady-states with the individual eigenstates $\ket{\psi}$ of the many-body Hamiltonian $\hat{H}$ for the free space case. As occurs in the two particle case, the steady-state has equal overlap with all $2^N$ eigenstates of the Hamiltonian for all simulatable particle numbers up to $N=8$. The steady-state emission rate, can be expressed as the expectation value of the dissipative Hamiltonian $\hat{H}_\mathrm{dis}=\sum_{i,j} \Gamma_{ij} \hat{\sigma}_i^+ \hat{\sigma}_j^-$. Then, each of the $2^N$ eigenstates of  $H_\mathrm{dis}$ can be assigned to one of the $N+1$ excitation subspaces containing $m \in \{0,N\}$ excitations. There are ${N \choose m} = M$ such states in the $m$-excitation subspace, each having a decay rate $\Gamma_i^{(m)}$, and such that the sum of all decay rates within the $m$-th excitation subspace is equal to $\sum_{i=1}^M \Gamma_i^{(m)}= m \gamma_0 {N \choose m}$. Since all eigenstates in all excitation manifolds are equally populated, the total emission rate is simply given as the average of all their decay rates
\begin{equation}
    \gamma_\mathrm{tot}^\mathrm{f.s.} = \sum_{m=0}^N \frac{\sum_{i=1}^M\Gamma_i^{(m)}}{2^N} = \frac{1}{2^N}\sum_{m=0}^N m \gamma_0 {N \choose m} = \frac{N}{2}\gamma_0.
    \label{eqn:freespace_decay_rate}
\end{equation}
This confirms the linear scaling of the total emission rate with particle number obtained in~\fref{fig:ss_scaling}(b). Intuitively, the fact that all eigenstates of $H_\mathrm{dis}$ are equally populated results in significant contribution of dark decay rates with $\Gamma_D \ll \gamma_0$ in the sum of~\eref{eqn:freespace_decay_rate}, which drastically diminishes the total emission rate.

In contrast, only the $N+1$ states contained in the symmetric subspace are occupied in the Dicke limit. The decay rate of the symmetric state in the $m$-excitation subspace, $| S=N/2,m \rangle$, is given as $\Gamma_m^\mathrm{Dicke} = \gamma_0[N/2 +(m-N/2)][N/2 - (m-N/2)+1]$, and the total emission rate then reads
\begin{align}
    \gamma_\mathrm{tot}^\mathrm{Dicke} &= \sum_{m=0}^N\frac{\Gamma_m^\mathrm{Dicke}}{N+1} = \frac{\gamma_0}{N+1}\sum_{m=0}^{N} m(N+1-m) \nonumber \\
    &= \gamma_0 \frac{N(N+2)}{6},
    \label{eqn:Dicke_decay_rate}
\end{align}
which highlights the quadratic scaling in particle number in the Dicke case. These findings are in agreement with the numerical results shown in~\ffref{fig:ss_scaling}(a) and (b). 

Hence, the stark difference in emission properties between free space and the paradigmatic Dicke limit can be explained by the contribution of the different dissipation channels available in the two cases.
While only radiative or superradiant states contribute to emission in the Dicke case, non-radiative dark or subradiant states are significantly occupied in the free space case, resulting in diminished photon emission.

\section{Finite Time Effects}
\label{sec:finite_time}
The emergence of decay channels with non-zero decay rates much smaller than $\gamma_0$ affect not only the steady-state emission properties of the system, as discussed in Sec.~\ref{sec:emission}, but also its dynamics. So far, we have characterized the steady-state properties by either determining the null-space of the Liouvillian or evolving approximate equations of motion for very long times ($t>100\gamma_0$). In experiments, however, the measurement of the emission properties typically takes place at much earlier times, and it is important to understand the interplay between the measurement time and the slowest or characteristic timescale at which the system evolves. To do so, we note that the time evolution of the density matrix $\hat{\rho}$ under the master equation~\eqref{eqn:EOM_densitymatrix} can be expressed as $\hat{\rho}(t) = \sum_{n=1}^{2^N}c_n e^{\lambda_n t} \mathbf{u}_n$, where the coefficients $c_n$ are fixed via the initial condition, and $\lambda_n$ and $\mathbf{u}_n$ respectively denote the eigenvalues and eigenvectors of the Liouvillian. Note that the eigenvalues are typically complex, $\lambda_n = -\kappa_n + i \nu_n$, and are characterized by a decay rate $\kappa_n$ and an energy shift $\nu_n$. Then, the fundamental timescale at which the steady-state is reached, $\tau_{ss}$, is equal to the inverse of the smallest non-zero decay rate of the Liouvillian spectrum with a non-zero contribution $c_n$.  In~\fref{fig:explan_fig}(c), we show the decay rates of the Lindbladian spectrum for two atoms, $\kappa_n = - \textit{Re} (\lambda_n)$. The smallest non-zero decay rate is indicated by yellow markers. For decreasing lattice spacing $a\rightarrow 0$, this eigenvalue approaches zero, which implies a divergence of the timescale required to reach the steady-state. This effect is nicely illustrated in~\fref{fig:explan_fig}(c), where we show the time evolution of the total emission rate for two atoms for different atom distances. For large enough lattice spacings, the steady-state emission rate reaches the analytic free space value given in~\eref{eqn:freespace_decay_rate} very quickly. For decreasing lattice spacings, the emission rate at finite times gets closer and closer to the Dicke value given in~\eref{eqn:Dicke_decay_rate}, as it takes longer and longer times for the subradiant states to be populated. For a lattice spacing of $a=0.05\lambda_0$ and after some oscillatory initial dynamics, the time evolution of the emission rate overlaps with that of the Dicke limit for the time window shown, $t \in \{ 0, 20\gamma_0 \}$. For such small lattice spacings, the subradiant decay rate is heavily suppressed and the time $\tau_{ss}$ required to reach the actual free space steady-state emission rate becomes increasingly large.

\begin{figure}
    \centering
    \includegraphics[width = 0.98\columnwidth]{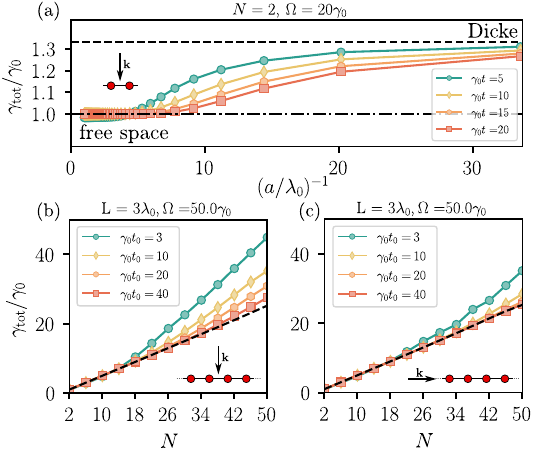}
    \caption{(a) Total emission rate as a function of inverse lattice spacing for two atoms.  For a measurement at finite times, the values for $\gamma_\mathrm{tot}$ approach the Dicke limit for decreasing density. (b) Scaling of the emission rate with particle number for a fixed chain length $L = 3\lambda_0$ and varying particle number, for different measurement times $t_0$ and plane wave drive perpendicular to the chain axis. The linear black dashed line correspond to measurement times $\gamma_0 t = 100$, for which the emission rate is consistent with the the steady-state analysis shown in~\fref{fig:ss_scaling}. For early measurement times, the scaling becomes superlinear and therefore resembles the Dicke case show in~\fref{fig:ss_scaling}(a). (c) Same as (b), but for plane-wave drive parallel to the chain axis. Compared to perpendicular drive, the emitted intensity is closer to the free space case and its characteristic linear scaling for all measurement times. (b) and (c) therefore show that the properties obtained for the two atom model in~\fref{fig:explan_fig}(e) also hold for larger particle numbers.}
    \label{fig:finite_time}
\end{figure}

In other words, the emergence of decay channels with heavily suppressed but non-zero decay rates increases the time required for the system to equilibrate and populate its dark states. This modifies the emission properties when measured at finite times (as it would occur in any experimental implementation). In~\fref{fig:finite_time}, we illustrate the effect of finite-time measurements on the total emission rate.
As shown in~\fref{fig:finite_time}(a), the emission rate approaches that of the Dicke limit for smaller interparticle spacings (\ie larger densities) and earlier measurement times.

Another important parameter which we did not consider thus far is the angle of the incident pump beam with respect to the orientation of the lattice. While the final steady-state is independent of the angle of incidence, the transient dynamics leading to this steady-state strongly depends on the phase factors $\propto e^{i\mathbf{k}\cdot\mathbf{r}_i}$ in~\eref{eq:Hamiltonian}. For the two atom case, we solve the full master equation [see~\fref{fig:explan_fig}(e)] and find that the effect of finite measurement times is most pronounced for perpendicular illumination ($\theta = \pi/2$), where the total emission rate gets close to the actual Dicke value at early times. For parallel illumination ($\theta = 0.0$), this effect is less pronounced, and we obtain emission rates much closer to the free space case for all measurement times. This phenomenon can be understood by noting the fundamental difference between both scenarios. For perpendicular illumination, the driving field couples the ground state $\ket{G}$ only to the bright state $\ket{B}$, and contributes to a rapid population of the superradiant states of the system. The subradiant state $\ket{D}$, however, only gets populated via the subradiant decay channel at a rate proportional to the slow timescale $\Gamma_D = \gamma_0 - \Gamma_{12}$. For parallel illumination, however, the driving field generally couples the ground state $\ket{G}$ to both the superradiant $\ket{B}$ and subradiant $\ket{D}$ single-excitation states, and the population of the latter can occur at a much faster timescale. As a result, the steady-state is also reached at earlier times, and the effect of measuring at finite times is substantially suppressed.

We further study the scaling of the emission rate as a function of density. To do so, we fix the length of the chain and analyze the total emission rate as a function of particle number $N$ for different measurement times [see~\fref{fig:finite_time}(b)-(c)]. We again simulate the dynamics of the system by means of a second order cumulant expansion (see Appendix~\ref{app:cumulants}) given the large particle numbers considered. For sufficiently late measurement times ($\gamma_0 t_0 = 100$), we retrieve the expected linear scaling characteristics of the steady-state analyzed in Sec.~\ref{sec:emission}. For earlier measurement times, however, the total emission rate scales superlinearly with particle number, confirming that the effect observed in the minimal two atom model also occurs for large system sizes. Again, the direction of the plane-wave driving field has the same effect as in the two-atom case. The strongest deviation from linear scaling is found for perpendicular drive, which predominantly excites the superradiant modes of the system. For driving along the chain axis, the coupling to the most superradiant modes is reduced \cite{AnaAsenjo_2017_general} and deviations from linear scaling can only be observed at very short measurement times.

\section{Conclusions}
Motivated by the recent experimental progress in generating controllable subwavelength geometries of quantum emitters, we analyzed the role of geometry induced shifts and decay rates on the steady-state properties of driven atom arrays. We particularly focused on the transition between a magnetic and a paramagnetic (superradiant) regime, and characterized the emission properties for varying geometries. We find that the emergence of collective shifts and dissipations in structured arrays results in a stark difference compared to the simple Dicke regime, where equal all-to-all interactions with vanishing shifts are assumed. In particular, the population of dark states destroys the quadratic scaling with particle number of the steady-state emitted light intensity, characteristic of the Dicke limit. Instead, only a linear scaling is observed for extended geometries. At finite measurement times and for very dense arrangements of atoms, however, we approach the Dicke-like behavior due to the diverging time scales to reach the actual steady-state. These insights could directly contribute to the interpretation of recent experimental observations in a dense driven pencil-shaped cloud of atoms, which experimentally realized the driven superradiant phase transition in free space~\cite{ferioli_non-equilibrium_2023}.

In a broader setting, driven dissipative spin models are an exciting avenue for future research. The control over various atomic degrees of freedom enables novel protocols and phenomena based on the interplay between dissipation and coherent drive. Such protocols could allow the realization of novel states of matter in the steady-state, and provide alternative routes for dissipation assisted state preparation of light and matter. The latter could be achievable in generalized analog settings as considered in this work, as well as in fully programmable quantum devices, where Lindbladian terms can in principle be engineered by performing appropriate randomized measurements to target a particular steady-state.

\begin{acknowledgments}
\vspace{-0.4 cm}
The numerical results were obtained using the \textsc{Quantumoptics.jl} package~\cite{kramer_quantumopticsjl_2018}. O.R.B. acknowledges support from Fundación Mauricio y Carlota Botton and from Fundació Bancaria “la Caixa” (LCF/BQ/AA18/11680093). S.O. is supported by a postdoctoral fellowship of the Max Planck Harvard Research Center for Quantum Optics. SFY would like to acknowledge funding from NSF through the CUA PFC (PHY-2317134), the QSense QLCI and PHY-2207972.
\end{acknowledgments}


\onecolumngrid
\appendix
\section{Green's function}
\label{app:Greensfunction}
The Green's function for a point dipole which determines the interaction strength $J_{ij}$ as well as the collective dissipation $\Gamma_{ij}$ in Eqs.~\eqref{eq:Hamiltonian} and ~\eqref{eq: Linbladian} can be written in Cartesian coordinates as \cite{GreensFunction_Chew,GreensFunction_novotny_hecht_2006}
\begin{align}
G_{\alpha \beta}(\mathbf{r},\omega) &= \frac{e^{i k r}}{4\pi r} \left[ \left( 1 + \frac{i}{kr} - \frac{1}{(kr)^2} \right) \delta_{\alpha\beta} \right. \nonumber \\
  &+ \left. \left(-1 - \frac{3i}{kr} + \frac{3}{(kr)^2} \right) \frac{r_\alpha r_\beta}{r^2} \right] + \frac{\delta_{\alpha \beta} \delta^{(3)}(\mathbf{r})}{3k^2},
\end{align}
where $k=\omega/c$, $r=|\mathbf{r}|$, and $\alpha,\beta=x,y,z$.

\section{Critical driving strength in the Dicke limit}
\label{app:Dicke_thres}
The mean-field equations in the Dicke limit are given as
\begin{subequations}
\begin{align}
\dot{s^x} &= \frac{1}{2}\gamma s^z s^x,\\
\dot{s^y} &= \frac{1}{2}\gamma s^z s^y - \Omega s^z,\\
\dot{s^z} &= -\frac{1}{2}\gamma \left[(s^x)^2 +  (s^y)^2\right] + \Omega s^y.
\end{align}
\label{eqn:Dicke_mf}
\end{subequations}
Solving for the steady-state by setting $\partial_t{s^{x,y,z}} = 0$ and imposing the additional constraint that $(s^x)^2 + (s^y)^2 + (s^z)^2 = N^2$ results in the stable mean-field steady-state solutions $s^x_\mathrm{MF} = 0$,
\begin{equation}
s^y_\mathrm{MF} =
\begin{cases}
\frac{\Omega}{\gamma} \quad \text{for} \, \, \Omega/N \in [0,1/2], \\
\frac{N^2\gamma}{4 \Omega} \quad \text{for} \, \, \Omega/N \in [1/2,1],
\end{cases}
\label{eqn:sy_Dicke_mf}
\end{equation}
and $s^z_\mathrm{MF} = -\frac{\sqrt{N^2 \gamma^2 - 4 \Omega^2}}{2 N \gamma}$. Theses solutions shown as solid black lines in~\fref{fig:mean-field_cuts}(a). Throughout this work we define the transition from the magnetized into the superradiant regime via the critical driving strength at which the spin $y$-component is maximal. Employing this condidtition we can directly read-off the critical Rabi frequency for the Dicke case from~\eqref{eqn:sy_Dicke_mf} to be $\Omega_\mathrm{mf}^\mathrm{Dicke} = N\gamma/2$.

\section{Linearized mean-field equations for the free space case}
\label{app:lin_eq}
To obtain a compact expression for the critical pump strength for different lattice spacings we linearize~\eeref{eqn:mf_simple} around a steady-state solution via the ansatz $s^{x,y,z}(t) = s_0^{x,y,z} + \delta s^{x,y,z}(t)$. Plugging this ansatz into~\eeref{eqn:mf_simple} and linearizing in $\delta s^{x,y,z}$ results in
\begin{subequations}
\begin{align}
\partial_t \delta s_x &= -\frac{\gamma}{2}s_0^x - \frac{\gamma}{2}\delta s^x + J s^y_0 s^z_0 + J s_y^0 \delta s_z + J s_z^0 \delta s_y + \frac{\Gamma}{2}s^x_0s^z_0 + \frac{\Gamma}{2} s^x_0 \delta s^z + \frac{\Gamma}{2}s^z_0 \delta s^x,\\
\partial_t \delta s_y &= -\frac{\gamma}{2}s_0^y - \frac{\gamma}{2}\delta s^y + J s^x_0 s^z_0 + J s_x^0 \delta s_z + J s_z^0 \delta s_x + \frac{\Gamma}{2}s^y_0s^z_0 + \frac{\Gamma}{2} s^y_0 \delta s^z + \frac{\Gamma}{2}s^z_0 \delta s^y - \Omega s^z_0 - \Omega \delta s_z,\\
\partial_t \delta s_z &= -\gamma\left(1 + s^z_0 +\delta s_z\right) - \frac{\Gamma}{2}\left[(s^0_x)^2 + s^x_0 \delta s_x + (s^0_y)^2 + s^y_0 \delta s_y\right] - \Omega s^y_0 + \Omega \delta s_y.
\end{align}
\label{eqn:lin_full}
\end{subequations}
For b rievity of notation we set $J_\mathrm{eff} \equiv J$ and $\Gamma_\mathrm{eff} \equiv J$.
Inserting the trivial steady-state solution for $\Omega = 0$,~\ie $s^x_0=s^y_0 = 0$, $s^z_0 = -1$ gives a linear set of equations
\begin{subequations}
\begin{align}
    \partial_t{\delta s^x} &= -\frac{\gamma}{2}\delta s^x - J \delta s^y - \frac{\Gamma}{2}\delta s^x, \\
    \partial_t{\delta s^y} &= -\frac{\gamma}{2}\delta s^y + J \delta s^x - \frac{\Gamma}{2}\delta s^y - \Omega \delta s^z + \Omega,\\
    \partial_t{\delta s^z} &= -\gamma\delta s^z + \Omega \delta s^y.
\end{align}
\label{eqn:lin_simple}
\end{subequations}
Solving these equations for the steady-state by putting $\partial_t \delta s^{x,y,z} = 0$ results in the steady-state solutions
\begin{subequations}
    \begin{align}
        \delta s^x_\mathrm{ss} &= -\frac{4 \gamma  J \Omega }{(\gamma +\Gamma ) \left(\gamma  (\gamma +\Gamma )+2 \Omega ^2\right)+4 \gamma  J^2},\\
        \delta s^y_\mathrm{ss} &= \frac{2 \gamma  \Omega  (\gamma +\Gamma )}{(\gamma +\Gamma ) \left(\gamma  (\gamma +\Gamma )+2 \Omega ^2\right)+4 \gamma  J^2},\\
        \delta s^z_\mathrm{ss} &= \frac{2 \Omega ^2 (\gamma +\Gamma )}{(\gamma +\Gamma ) \left(\gamma  (\gamma +\Gamma )+2 \Omega ^2\right)+4 \gamma  J^2}.
    \end{align}
\end{subequations}
The critical value for $\Omega$ is then determined by solving $\frac{d}{d\Omega}\delta s^y_\mathrm{ss} = 0$, which results in~\eqref{eqn:thres} in the main text.

\section{Second order cumulant expansion}
\label{app:cumulants}
Calculating the equations of motion of the $\hat{\sigma}_{x,y,z}$ operators via $\partial_t\langle\hat{O}\rangle = \Tr([\partial_t \hat{\rho}(t)] \hat{O})$, where $\hat{\rho}(t)$ is governed by~\eref{eqn:EOM_densitymatrix} results in the set of equations (for notational brevity we omit the hat symbol $\hat{\bullet}$ for operators below).

\begin{align}
\partial_t\langle{\sigma_k^x}\rangle &= \sum_{i;i \neq k} J_{ki} \langle\sigma_i^y\sigma_k^z\rangle -\frac{1}{2} \gamma \langle\sigma_k^x\rangle +\frac{1}{2} \sum_{i;i \neq k} \Gamma_{ki} \langle\sigma_i^x\sigma_k^z\rangle  - \Omega \sin(\mathbf{k} \cdot \mathbf{r}_k) \langle \sigma_k^z \rangle\\
\partial_t\langle{\sigma_k^y}\rangle &= -\sum_{i;i \neq k} J_{ki} \langle\sigma_i^x\sigma_k^z\rangle -\frac{1}{2} \gamma \langle\sigma_k^y\rangle +\frac{1}{2} \sum_{i;i \neq k} \Gamma_{ki} \langle\sigma_i^y\sigma_k^z\rangle  - \Omega \cos(\mathbf{k} \cdot \mathbf{r}_k) \langle \sigma_k^z \rangle\\
\partial_t\langle{\sigma_k^z}\rangle &= - \sum_{i;i \neq k} J_{ki} \Big(\langle\sigma_k^x\sigma_i^y\rangle - \langle\sigma_i^x\sigma_k^y\rangle\Big) -\gamma \big(1 + \langle\sigma_k^z\rangle\big) -\frac{1}{2} \sum_{i;i \neq k} \Gamma_{ki} \Big(\langle\sigma_k^x\sigma_i^x\rangle + \langle\sigma_i^y\sigma_k^y\rangle\Big)\\
& \quad + \Omega \sin(\mathbf{k} \cdot \mathbf{r}_k) \langle \sigma_k^x \rangle + \Omega \cos(\mathbf{k} \cdot \mathbf{r}_k) \langle \sigma_k^y \rangle
\label{eqn:first_oder_no_mf}
\end{align}

Calculating additional equations of motion for the two-point correlators $\propto \langle O_i O_k\rangle$ and replacing averages over third-order operators by \cite{Cumulant_Kubo, Ritsch_cumulants_package} 
\begin{align}
\langle O_1 O_2 O_3 \rangle &= \langle O_1 \rangle \langle O_2 O_3 \rangle + \langle O_2 \rangle \langle O_1 O_3 \rangle \nonumber \\
&+ \langle O_3 \rangle \langle O_1 O_2 \rangle -2 \langle O_1 \rangle \langle O_2 \rangle \langle O_3 \rangle,
\end{align}
results in a closed set of differential equations. The equations for the correlators are given as
\begin{align}
\partial_t\langle{\sigma_k^x\sigma_l^x}\rangle &= \sum_{j;j \neq k,l} J_{kj} \langle\sigma_k^z\sigma_l^x\sigma_j^y\rangle + \sum_{j;j \neq k,l} J_{lj} \langle\sigma_k^x\sigma_l^z\sigma_j^y\rangle  - \gamma \langle\sigma_k^x\sigma_l^x\rangle + \Gamma_{kl} \Big( \langle\sigma_k^z\sigma_l^z\rangle + \frac{1}{2} \langle\sigma_k^z\rangle + \frac{1}{2} \langle\sigma_l^z\rangle \Big) \nonumber \\&\quad + \frac{1}{2} \sum_{j;j \neq k,l} \Gamma_{kj} \langle\sigma_k^z\sigma_l^x\sigma_j^x\rangle + \frac{1}{2} \sum_{j;j \neq k,l} \Gamma_{lj} \langle\sigma_k^x\sigma_l^z\sigma_j^x\rangle - \Omega \Big(\sin(\mathbf{k} \cdot \mathbf{r}_l)\langle\sigma_k^x\sigma_l^z\rangle + \sin(\mathbf{k} \cdot \mathbf{r}_k)\langle\sigma_l^x\sigma_k^z\rangle\Big)\\
\partial_t\langle{\sigma_k^y\sigma_l^y}\rangle &= - \sum_{j;j \neq k,l} J_{kj} \langle\sigma_k^z\sigma_l^y\sigma_j^x\rangle - \sum_{j;j \neq k,l} J_{lj} \langle\sigma_k^y\sigma_l^z\sigma_j^x\rangle - \gamma \langle\sigma_k^y\sigma_l^y\rangle + \Gamma_{kl}\Big( \langle\sigma_k^z\sigma_l^z\rangle +\frac{1}{2} \langle\sigma_k^z\rangle +\frac{1}{2} \langle\sigma_l^z\rangle \Big)  \nonumber \\&\quad +\frac{1}{2} \sum_{j;j \neq k,l} \Gamma_{kj} \langle\sigma_k^z\sigma_l^y\sigma_j^y\rangle +\frac{1}{2} \sum_{j;j \neq k,l} \Gamma_{lj} \langle\sigma_k^y\sigma_l^z\sigma_j^y\rangle - \Omega\Big(\cos(\mathbf{k} \cdot \mathbf{r}_l)\langle\sigma_k^y\sigma_l^z\rangle + \cos(\mathbf{k} \cdot \mathbf{r}_k)\langle\sigma_l^y\sigma_k^z\rangle\Big)\\
\partial_t\langle \sigma_k^z\sigma_l^z \rangle &= \sum_{j;j \neq k,l} J_{kj} \Big( \langle \sigma_k^y\sigma_l^z\sigma_j^x \rangle - \langle \sigma_k^x\sigma_l^z\sigma_j^y \rangle \Big) +\sum_{j;j \neq k,l} J_{lj} \Big( \langle \sigma_k^z\sigma_l^y\sigma_j^x \rangle - \langle \sigma_k^z\sigma_l^x\sigma_j^y \rangle \Big) - 2 \gamma \langle \sigma_k^z\sigma_l^z \rangle - \gamma \big( \langle \sigma_l^z \rangle + \langle \sigma_k^z \rangle \big) \nonumber \\
&\qquad +\Gamma_{kl}\Big( \langle \sigma_k^y\sigma_l^y \rangle + \langle \sigma_k^x\sigma_l^x \rangle \Big) -\frac{1}{2} \sum_{j;j \neq k,l} \Gamma_{kj} \Big( \langle \sigma_k^x\sigma_l^z\sigma_j^x \rangle + \langle \sigma_k^y\sigma_l^z\sigma_j^y \rangle \Big) \nonumber \\
&\qquad-\frac{1}{2} \sum_{j;j \neq k,l} \Gamma_{lj} \Big( \langle \sigma_k^z\sigma_l^x\sigma_j^x \rangle + \langle \sigma_k^z\sigma_l^y\sigma_j^y \rangle \Big) + \Omega\Big(\cos(\mathbf{k} \cdot \mathbf{r}_k) \langle \sigma_k^y\sigma_l^z \rangle + \cos(\mathbf{k} \cdot \mathbf{r}_l) \langle \sigma_l^y\sigma_k^z \rangle \Big) \nonumber\\
&\qquad +\Omega\Big(\sin(\mathbf{k} \cdot \mathbf{r}_k) \langle \sigma_k^x\sigma_l^z \rangle + \sin(\mathbf{k} \cdot \mathbf{r}_l) \langle \sigma_l^x\sigma_k^z \rangle \Big)\\
\partial_t\langle{\sigma_k^x\sigma_l^y}\rangle &= J_{kl}\Big( \langle\sigma_k^z\rangle - \langle\sigma_l^z\rangle \Big) +\sum_{j;j \neq k,l} J_{kj} \langle\sigma_k^z\sigma_l^y\sigma_j^y\rangle -\sum_{j;j \neq k,l} J_{lj} \langle\sigma_k^x\sigma_l^z\sigma_j^x\rangle - \gamma \langle\sigma_k^x\sigma_l^y\rangle + \frac{1}{2} \sum_{j;j \neq k,l} \Gamma_{kj} \langle\sigma_k^z\sigma_l^y\sigma_j^x\rangle \nonumber \\&\quad + \frac{1}{2} \sum_{j;j \neq k,l} \Gamma_{lj} \langle\sigma_k^x\sigma_l^z\sigma_j^y\rangle - \Omega\cos(\mathbf{k} \cdot \mathbf{r}_l)\langle\sigma_k^x\sigma_l^z\rangle - \Omega\sin(\mathbf{k} \cdot \mathbf{r}_k)\langle\sigma_l^y\sigma_k^z\rangle\\
\partial_t\langle{\sigma_k^x\sigma_l^z}\rangle &= J_{kl} \langle\sigma_l^y\rangle +\sum_{j;j \neq k,l} J_{kj} \langle\sigma_k^z\sigma_l^z\sigma_j^y\rangle +\sum_{j;j \neq k,l} J_{lj} \Big( \langle\sigma_k^x\sigma_l^y\sigma_j^x\rangle -\langle\sigma_k^x\sigma_l^x\sigma_j^y\rangle \Big) - \frac{3}{2} \gamma \langle\sigma_k^x\sigma_l^z\rangle - \gamma \langle\sigma_k^x\rangle \nonumber \\&\quad - \Gamma_{kl}\Big( \langle\sigma_k^z\sigma_l^x\rangle +\frac{1}{2} \langle\sigma_l^x\rangle \Big) + \frac{1}{2} \sum_{j;j \neq k,l} \Gamma_{kj} \langle\sigma_k^z\sigma_l^z\sigma_j^x\rangle - \frac{1}{2} \sum_{j;j \neq k,l} \Gamma_{lj} \Big( \langle\sigma_k^x\sigma_l^x\sigma_j^x\rangle +\langle\sigma_k^x\sigma_l^y\sigma_j^y\rangle \Big) + \Omega\cos(\mathbf{k} \cdot \mathbf{r}_l)\langle\sigma_k^x\sigma_l^y\rangle \nonumber\\ &\qquad + \Omega \Big(\sin(\mathbf{k} \cdot \mathbf{r}_l)\langle\sigma_k^x \sigma_l^x \rangle - \sin(\mathbf{k} \cdot \mathbf{r}_k)\langle \sigma_k^z \sigma_l^z \rangle\Big)\\
\partial_t\langle{\sigma_k^y\sigma_l^z}\rangle &= -J_{kl} \langle\sigma_l^x\rangle -\sum_{j;j \neq k,l} J_{kj} \langle\sigma_k^z\sigma_l^z\sigma_j^x\rangle +\sum_{j;j \neq k,l} J_{lj} \Big( \langle\sigma_k^y\sigma_l^y\sigma_j^x\rangle -\langle\sigma_k^y\sigma_l^x\sigma_j^y\rangle \Big) - \frac{3}{2} \gamma \langle\sigma_k^y\sigma_l^z\rangle - \gamma \langle\sigma_k^y\rangle \nonumber \nonumber \\&\quad - \Gamma_{kl}\Big( \langle\sigma_k^z\sigma_l^y\rangle + \frac{1}{2}\langle\sigma_l^y\rangle \Big) + \frac{1}{2} \sum_{j;j \neq k,l} \Gamma_{kj} \langle\sigma_k^z\sigma_l^z\sigma_j^y\rangle - \frac{1}{2} \sum_{j;j \neq k,l} \Gamma_{lj} \Big( \langle\sigma_k^y\sigma_l^x\sigma_j^x\rangle +\langle\sigma_k^y\sigma_l^y\sigma_j^y\rangle \Big) + \Omega\sin(\mathbf{k} \cdot \mathbf{r}_l)\langle\sigma_l^x\sigma_k^y\rangle \nonumber \\ &\qquad + \Omega \Big( \cos(\mathbf{k} \cdot \mathbf{r}_l)\langle\sigma_k^y \sigma_l^y \rangle - \cos(\mathbf{k} \cdot \mathbf{r}_k)\langle \sigma_k^z \sigma_l^z \rangle\Big)
\label{eqn:secondorder}
\end{align}
The truncation in the considered correlations allows the simulation of larger particle numbers compared to the free space case. These equations are used to determine the scaling of the total emission rate for the free space case obtained in~\ref{sec:emission}. It is given as
\begin{equation}
    \gamma_\mathrm{tot}(t) = \gamma_0 \sum_{k=1}^N\frac{\langle\sigma_k^z\rangle(t) + 1}{2} + \sum_{k,l=1}^N \Gamma_{kl}\frac{\langle\sigma_k^x \sigma_l^x\rangle(t) + \langle\sigma_k^y \sigma_l^y\rangle(t)}{4}.
    \label{eqn:tot_emission_rate_cumulants}
\end{equation}
The steady-state value can be obtained by evolving~\eeref{eqn:first_oder_no_mf} and~\eqref{eqn:secondorder} for long times $t_\mathrm{max} \gg 1/\gamma_0$ and evaluating~\eref{eqn:tot_emission_rate_cumulants} at the final time.

\section{Analytical steady-state solutions for the two atom case}
\label{app:two-atoms}

$\hat{H}_\mathrm{drive} = \Omega\left(e^{-i k d/2} \hat{\sigma}^-_1 + e^{i k d/2} \hat{\sigma}^-_2 + \text{H.c.} \right) = \Omega \cos(kd/2)(\hat{\sigma}_1 + \hat{\sigma}_2) - i\Omega \sin(kd/2)(\hat{\sigma}_1-\hat{\sigma}_2) + \text{H.c.}$
For the two atom case the system Hamiltonian $\hat{H}$ can be decomposed into the eigenstates $\ket{G} = \ket{gg}$, $\ket{B} = 1/\sqrt{2}(\ket{eg} + \ket{ge})$, $\ket{D} = 1/\sqrt{2}(\ket{eg} - \ket{ge})$ and $\ket{E} = \ket{ee}$ as
\begin{align}
    \hat{H}_\mathrm{int} &= - 2 \Delta \ket{E}\bra{E} + (-\Delta + J)\ket{B}\bra{B} - (\Delta+J)\ket{D}\bra{D}\\
    \hat{H}_\mathrm{drive} &= \Omega_B \left(\ket{B}\bra{G} + \ket{G}\bra{B} + \ket{B}\bra{E} + \ket{E}\bra{B}\right)\\
    &\quad + i \Omega_D \left(\ket{D}\bra{G} - \ket{G}\bra{D} + \ket{D}\bra{E} - \ket{E}\bra{D}\right),
\end{align}
with $\Omega_B \equiv \sqrt{2}\Omega \cos(kd/2)$, $\Omega_D \equiv \sqrt{2} \Omega \sin(kd/2)$,  $d$ denotes the distance between the two atoms and we introduced the detuning $\Delta = \omega - \omega_0$. Equivalently the Liouvillian can be written as
\begin{align}
    \mathcal{L} &= \frac{\Gamma_B}{2}\left(\ket{B}\bra{B}\hat{\rho} + \hat{\rho} \ket{B}\bra{B}\right) - \frac{\Gamma_D}{2}\left(\ket{D}\bra{D}\hat{\rho} + \hat{\rho} \ket{D}\bra{D}\right) - \frac{\Gamma_B + \Gamma_D}{2}\left(\ket{E}\bra{E}\hat{\rho} + \hat{\rho} \ket{E}\bra{E}\right)\\
    & \quad +\Gamma_B \left(\bra{B}\hat{\rho}\ket{B}\ket{G}\bra{G} + \bra{E}\hat{\rho}\ket{E}\ket{B}\bra{B} + \bra{B}\hat{\rho}\ket{E}\ket{G}\bra{B} + \bra{E}\hat{\rho}\ket{B}\braket{B|G}\right)\\
    & \quad +\Gamma_D \left(\bra{D}\hat{\rho}\ket{D}\ket{G}\bra{G} + \bra{E}\hat{\rho}\ket{E}\ket{D}\bra{D} - \bra{D}\hat{\rho}\ket{E}\ket{G}\bra{D} + \bra{E}\hat{\rho}\ket{D}\braket{D|G}\right),
\end{align}
with the bright (dark) state decay rate $\Gamma_B$ ($\Gamma_D$).

In this dressed state picture the dynamics of the state populations $p_G = \bra{G}\hat{\rho}\ket{G}$, $p_E = \bra{E}\hat{\rho}\ket{E}$, $p_B = \bra{B}\hat{\rho}\ket{B}$ and $p_D = \bra{D}\hat{\rho}\ket{D}$, as well as the correlators $c_{EB} = \bra{E}\hat{\rho}\ket{B}$, $c_{ED} = \bra{E}\hat{\rho}\ket{D}$, $c_{EG} = \bra{E}\hat{\rho}\ket{G}$, $C_{BD} = \bra{B}\hat{\rho}\ket{D}$, $c_{BG} = \bra{B}\hat{\rho}\ket{G}$, $c_{DG} = \bra{D}\hat{\rho}\ket{G}$ is governed by the master equation~\eqref{eqn:EOM_densitymatrix}. It is given as
\begin{subequations}
\begin{align}
    \dot{p_E} &= -(\Gamma_B + \Gamma_D)p_E - i \Omega_B (c_{BE} - c_{EB}) - \Omega_D (c_{DE} + c_{ED}),\\
    \dot{p_B} &= -\Gamma_B p_B + \Gamma_B p_E + i \Omega_B (c_{BE} - c_{EB}) - i\Omega_B (c_{GB} - c_{BG}),\\
    \dot{p_D} &= -\Gamma_D p_D + \Gamma_D p_E + \Omega_D (c_{DE} - c_{ED}) + \Omega_D (c_{GD} - c_{DG}),\\
    \dot{p_G} &= \Gamma_B p_B + \Gamma_D p_D + i \Omega_B (c_{GB} - c_{BG}) - \Omega_D (c_{GD} + c_{DG}),\\
    \dot{c_{EB}} &= -\frac{2\Gamma_B + \Gamma_D}{2}c_{EB} + i(J + \Delta)c_{EB} - i \Omega_B(p_B-p_E) + i\Omega c_{EG}-\Omega_D C_{DB}\\
    \dot{c_{ED}} &= -\frac{2\Gamma_D + \Gamma_B}{2}c_{ED} - i(J - \Delta)c_{ED} - i \Omega_B c_{BD} - \Omega_D (p_B-p_D) - \Omega_D C_{EG}\\
    \dot{c_{EG}} &= -\frac{\Gamma_B + \Gamma_D}{2}c_{EG} + 2i\Delta c_{EG} -i\Omega_B(c_{BG} - c_{EB}) - \Omega_D(c_{DG} + c_{ED}),\\
    \dot{c_{BD}} &= -\frac{\Gamma_B + \Gamma_D}{2}c_{BD} - 2iJ c_{BD} -i\Omega_B(c_{ED} + c_{GD}) + \Omega_D(c_{BE} + c_{BG})\\
    \dot{c_{BG}} &= -\frac{\Gamma_B}{2}c_{BD} + \Gamma_B c_{EB} -i(J-\Delta)c_{BG} +i\Omega_B(p_B - p_G) - i\Omega_B c_{EG} - \Omega_D c_{BD}\\
    \dot{c_{DG}} &= -\frac{\Gamma_D}{2}c_{DG} + \Gamma_{DC} c_{ED} + i(J+\Delta)c_{DG} - \Omega_D(p_D - p_G) + \Omega_D c_{EG} + i \Omega_B c_{DB}
    \end{align}
\end{subequations}
Since classical drive can only drive the symmetric bright state $\Omega_D = 0$. Solving these equations under the assumption of resonant drive of the bright state,~\ie $\Delta = J$, results in the steady-state solutions in the Dicke limit ($\Gamma_D = 0$ and $J=0$)

\begin{subequations}
\begin{align}
    p_E &= \frac{\Omega_B^4}{(\Omega_B^2 + \left(\frac{\Gamma_B}{2}\right)^2)^2 + 2 \Omega_B^4},\\
    p_B &= \frac{\Omega_B^2(\Omega_B^2 + \left(\frac{\Gamma_B}{2}\right)^2)}{(\Omega_B^2 + \left(\frac{\Gamma_B}{2}\right)^2)^2 + 2 \Omega_B^4},\\
    p_D &= 0,\\
    p_G &= 1 - p_E - p_B
\end{align}
\end{subequations}
For $\Omega_B \gg \Gamma_B$ these solutions converge to $p_E = p_B = p_G = 1/3$. The total emission rate in this case is given as the sum of the populations in $\ket{E}$ and $\ket{B}$ and amounts to $\gamma_\mathrm{tot}^\mathrm{Dicke} = \Gamma_B (p_B + p_E) = 2\gamma_0 (p_B + p_E) = 4/3\gamma_0$.

In the general free space case ($J,\Gamma_D\neq 0$) the steady-state solutions are
\begin{subequations}
\label{eq:app_pops_general}
\begin{align}
    p_E=p_D &= \frac{\Omega_B^4}{4\Omega_B^4 + \left(2\Omega_B^2 + \left(\frac{\Gamma_B}{2}\right)\right)\left(4J^2 + \left(4J^2 + \left(\frac{\Gamma_B + \Gamma_D}{2}\right)^2\right)\right)},\\
    p_B &= \frac{\Omega_B^4 + \Omega_B^2 \left(4J^2 + \left(\frac{\Gamma_B + \Gamma_D}{2}\right)^2\right)}{4\Omega_B^4 + \left(2\Omega_B^2 + \left(\frac{\Gamma_B}{2}\right)\right)\left(4J^2 + \left(4J^2 + \left(\frac{\Gamma_B + \Gamma_D}{2}\right)^2\right)\right)},\\
    p_G &= 1 - p_E - p_B - p_D,
\end{align}
\end{subequations}
which in the $\Omega \gg \Gamma_B$ limit results in equal population of each eigenstates in the steady-state ($p_G = p_E = p_D = p_B = 1/4$). Note that, while we have assumed for simplicity $\Omega_D=0$ in Eq.~(\ref{eq:app_pops_general}), the same steady-state would be found for non-zero and large $\Omega_D$. That is, in the general free space case, the total emission rate is given $\gamma_\mathrm{tot}^\mathrm{f.s} =(\Gamma_B+\Gamma_D)p_E + \Gamma_B p_B + \Gamma_D p_D = \gamma_0$. It is this fundamental difference in the steady-state populations between the Dicke and free space case, in particular the population of the dark state in the free space case, which gives provides an explanation for the stark difference in emission properties for the two cases as discussed in the main text.
 
\twocolumngrid
\bibliography{reference_driven_array}
\end{document}